\documentclass[conference]{IEEEtran}
\IEEEoverridecommandlockouts
\usepackage{cite}
\usepackage{amsmath,amssymb,amsfonts}
\usepackage{algorithmic}
\usepackage{graphicx}
\usepackage{textcomp}
\usepackage{xcolor}
\usepackage{hyperref}
\usepackage{enumitem}
\usepackage{subcaption}
\usepackage{mdframed}
\usepackage{caption}
\mdfdefinestyle{graybox}{
    backgroundcolor=gray!20,
    linecolor=black,
    linewidth=1pt,
    roundcorner=5pt,
    skipabove=10pt,
    skipbelow=10pt,
    innertopmargin=6pt,
    innerbottommargin=6pt,
    innerleftmargin=6pt,
    innerrightmargin=6pt
}
\usepackage{url}
\def\BibTeX{{\rm B\kern-.05em{\sc i\kern-.025em b}\kern-.08em
    T\kern-.1667em\lower.7ex\hbox{E}\kern-.125emX}}
\begin{document}

\title{Automated Extract Method Refactoring with Open-Source LLMs: A Comparative Study\\

}

\author{
  \IEEEauthorblockN{
    Sivajeet Chand\IEEEauthorrefmark{1}\quad
    Melih Kilic\IEEEauthorrefmark{1}\quad
    Roland Würsching\IEEEauthorrefmark{1}\quad
    Sushant Kumar Pandey\IEEEauthorrefmark{2}\quad
    Alexander Pretschner\IEEEauthorrefmark{1}
  }
  \IEEEauthorblockA{\IEEEauthorrefmark{1}Technical University of Munich, Munich, Germany\\
  \texttt{\{sivajeet.chand, roland.wuersching, alexander.pretschner\}@tum.de, ge85zak@mytum.de}}
 % \IEEEauthorblockA{\IEEEauthorrefmark{2}Technical University of Munich, Munich, Germany\\
  %\texttt{ge85zak@mytum.de}}
  \IEEEauthorblockA{\IEEEauthorrefmark{3}University of Groningen, Groningen, The Netherlands\\
  \texttt{s.k.pandey@rug.nl}}
}

\maketitle

\begin{abstract}
Automating the Extract Method refactoring (EMR) remains challenging and largely manual despite its importance in improving code readability and maintainability. Recent advances in open-source, resource-efficient Large Language Models (LLMs) offer promising new approaches for automating such high-level tasks. In this work, we critically evaluate five state-of-the-art open-source LLMs, spanning 3B to 8B parameter sizes, on the EMR task for Python code. We systematically assess functional correctness and code quality using automated metrics and investigate the impact of prompting strategies by comparing one-shot prompting to a Recursive criticism and improvement (RCI) approach. RCI-based prompting consistently outperforms one-shot prompting in test pass rates and refactoring quality. The best-performing models, Deepseek-Coder-RCI and Qwen2.5-Coder-RCI, achieve test pass percentage (TPP) scores of 0.829 and 0.808, while reducing lines of code (LOC) per method from 12.103 to 6.192 and 5.577, and cyclomatic complexity (CC) from 4.602 to 3.453 and 3.294, respectively. A developer survey on RCI-generated refactorings shows over 70\% acceptance, with Qwen2.5-Coder rated highest across all evaluation criteria. In contrast, the original code scored below neutral, particularly in readability and maintainability, underscoring the benefits of automated refactoring guided by quality prompts. While traditional metrics like CC and LOC provide useful signals, they often diverge from human judgments, emphasizing the need for human-in-the-loop evaluation. Our open-source benchmark offers a foundation for future research on automated refactoring with LLMs.
\end{abstract}


\section{Introduction}
Automated code refactoring is a foundational Software Engineering (SE) practice that improves code readability\cite{Inoue2017MORE:}, maintainability\cite{Meananeatra2012Identifying}, and long-term evolution \cite{fowler2018refactoring, kim2012refactoring}. Among refactoring techniques, the Extract Method is one of the most frequently applied in both industry and open-source development and impactful, enabling developers to modularize complex code, reduce duplication, and clarify program structure \cite{fowler2018refactoring, agnihotri2020systematic}. This technique, whereby developers decompose long or complex functions into smaller, well-named methods, directly targets the ``long method" code smell, a persistent source of defects and maintenance challenges\cite{agnihotri2020systematic, lacerda2020code}. By isolating logically coherent fragments of code, the Extract Method improves modularity, supports better testing, and simplifies future code evolution. However, despite its importance, the EMR process remains manual, mainly reliant on rule-based tools with limited flexibility and adaptability \cite{moha2010decor, alikacem2009metric, lacerda2020code}.

The recent rise of LLMs, particularly those capable of code understanding and generation, has catalyzed new research in automated code transformation \cite{li2023starcoder, chen2021evaluating, roziere2023code}. While proprietary LLMs such as OpenAI's Codex and GPT-4 have shown promise in code-related tasks \cite{chen2021evaluating, achiam2023gpt4, pandey2025design}, the rapid advancement of open-source LLMs has made high-quality, accessible, and transparent models available to both researchers and practitioners \cite{abouelenin2025phi4mini, grattafiori2024llama3, hui2024qwen25, guo2024deepseekcoder}. The ability of open-source models to automate nuanced software engineering tasks like EMR remains largely unevaluated. Unlike code synthesis, EMR demands that LLMs understand existing code, extract cohesive fragments, preserve functionality, and generate clear method names consistent with the codebase. Another open question is how best to interact with LLMs for these tasks. Prompt engineering is now recognized as a crucial determinant of LLM performance\cite{kim2023language}, yet the specific impact of advanced prompting strategies, such as recursive criticism and improvement (RCI), on automated refactoring outcomes remains underexplored.

The SE community traditionally evaluates refactoring through quantitative metrics, such as cyclomatic complexity (CC)\cite{Alcocer2021Quality, 2023The, Hummel2016Deriving, Nasrabadi2022An} and lines of code (LOC)\cite{Morishita2023Refactoring, Hora2020Characteristics, Hummel2016Deriving, Nasrabadi2022An}. However, it is increasingly evident that these metrics do not always reflect what human developers perceive as high-quality or acceptable refactoring\cite{shirafuji2023refactoring, eilertsen2020refactoring, ikhsan2018auto}. This disconnect raises essential questions about properly evaluating automated refactoring tools, especially those powered by LLMs.

This paper addresses this specific gap by providing a systematic, human-centered understanding of the extent to which open-source LLMs can perform EMR. We compare five state-of-the-art open-source LLMs, focusing on efficient models to run on commonly available hardware\cite{guo2024deepseekcoder, abdin2024phi4, achiam2023gpt4, grattafiori2024llama3}. We investigate the impact of the prompting strategy, contrasting one-shot prompting with an RCI method \cite{kim2023language}, and evaluate the results using quantitative code metrics and a qualitative developer survey.

The models that demonstrate the best performance are Deepseek-Coder-RCI and Qwen2.5-Coder-RCI. These models achieve EM TPP scores of 0.829 and 0.808, respectively, while concomitantly reducing code length from 12.103 LOC to 6.192 and 5.577, and code complexity from 4.602 CC to 3.453 and 3.294, respectively. Our findings reveal that the prompting strategy substantially impacts refactoring success and code quality and that traditional metrics often diverge from human acceptance of refactoring outcomes. Importantly, we demonstrate that when properly prompted, open-source LLMs can produce EMRs that developers accept in over 70\% of cases, signaling their readiness for real-world adoption.

By releasing our benchmarking pipeline, we aim to foster reproducible, community-driven progress in automated refactoring research and practice.\footnote{\url{https://figshare.com/s/8f242ba2a071198eb4eb}} Our key Contributions include:

\textbf{\textit{1) Benchmarking Open-Source LLMs:}} First systematic evaluation of five 3B–8B open-source models on the EMR task for Python.

\textbf{\textit{2) Prompting Strategies:}} Comparison of one-shot vs. RCI, showing RCI significantly boosts refactoring success.

\textbf{3) \textit{Evaluation Methods:}} Combined automated metrics (e.g., TPP, CC, LOC) with developer survey, revealing divergences between metrics and human judgment.

\textbf{4) \textit{Practical Viability:}} Demonstrated that properly prompted open-source models achieve $>70\%$ developer acceptance on commodity hardware.

\section{Related Work}

\textbf{\textit{Traditional EMR Tools:}} EMR has long been used to improve code clarity, reduce duplication, and enhance maintainability. Traditional EMR tools are rule-based, using deterministic patterns and syntactic heuristics to extract cohesive code blocks into methods.

Early work, such as \cite{cortes2006rule}, laid the foundation for rule-based EMR by applying predefined transformation templates to support framework evolution and component reuse. Mainstream tools like Eclipse JDT, IntelliJ IDEA, and Ref-Finder continue to offer EMR capabilities based on static code analysis. However, as \cite{khaleel2024automatic} observed, these tools often struggle with semantically complex or context-sensitive code segments. To address modularity in legacy systems, \cite{marcos2009aspect} combined aspect mining with rule-based EMR, enabling the clean separation of cross-cutting concerns. Similarly, \cite{ghannem2018model} integrated EMR into a multi-objective rule-based framework to support higher-level design transformations. \cite{sharma2021explainable} incorporated transparent decision paths for rule-based EMR suggestions, enhancing developer trust and tool adoption.

\cite{alharbi2024comparative} classified EMR techniques, noting that rule-based variants are reliable in static scenarios but inflexible in dynamic or behavior-driven contexts. In contrast, \cite{alOmar2023just} proposed a just-in-time, clone-aware EMR tool that operates continuously during development. \cite{baqais2020automatic} described rule-based EMR strategies as low-risk and predictable, making them particularly valuable in industrial software maintenance. However, they noted that such approaches trade adaptability for reliability compared to modern AI-driven alternatives.

\textbf{\textit{Large Language Models for Automated Refactoring:}} The emergence of LLMs trained on massive code corpora is shifting code refactoring into a data‑driven, AI‑powered process. Early models such as OpenAI’s Codex\cite{openai2021codex} and, StarCoder\cite{li2023starcoder}, followed by open-source alternatives like Code Llama\cite{grattafiori2024llama3}, DeepSeek-Coder\cite{guo2024deepseekcoder}, and Qwen2.5-Coder\cite{hui2024qwen25}, have demonstrated impressive capabilities in code completion and generation. Standard benchmarks like HumanEval\cite{chen2021evaluating} and EvalPlus\cite{liu2023is} have accelerated comparative research in this domain.

Recent work has begun to investigate the use of LLMs for automated refactoring. Shirafuji et al.\cite{shirafuji2023refactoring} explored few-shot prompting with GPT-3.5 for various refactoring tasks, achieving measurable code size and complexity reductions. Choi et al.\cite{choi2024iterative} extended this idea, proposing iterative LLM-based refactoring strategies combined with regression testing for validation. Pomian et al.\cite{pomian2024together} have shown that LLMs can support EMR, using GPT-3.5, integrated directly into development environments for real-time developer assistance. In contrast, our study focuses on benchmarking open-source, resource-efficient LLMs (3B–8B) on commodity hardware, relying exclusively on prompting-only strategies. Rather than hybridizing LLMs with auxiliary tools or retrieval components, we aim to identify which small-scale model performs best under constrained settings. Notably, the best-performing model from our benchmark could serve as a foundation for future hybrid approaches, including those proposed by Pomian et al.

Despite these advances, most existing studies focus on proprietary or large-scale LLMs (e.g., GPT-3.5, GPT-4), which may be inaccessible or impractical for many organizations. Furthermore, the impact of prompt engineering, particularly advanced strategies like RCI\cite{kim2012refactoring}, on LLM-driven refactoring outcomes remains underexplored. 

Our work addresses these gaps by systematically benchmarking state-of-the-art open-source LLMs on the EMR task, rigorously comparing prompting techniques, and incorporating automated metrics and human developer feedback.

\section{Methodology}
\begin{figure}[ht!]
    \centering
    \includegraphics[width=9cm]{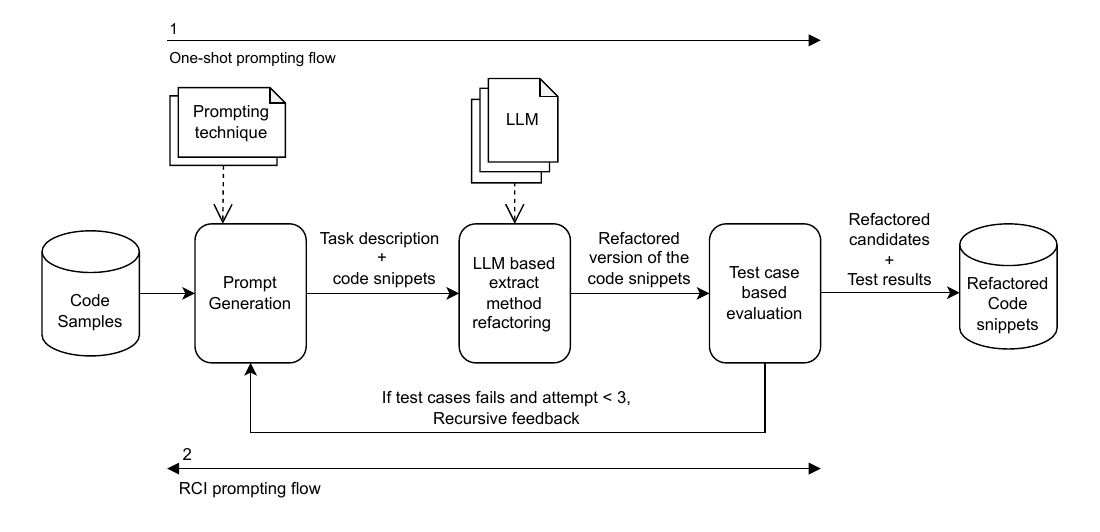}
    \caption{Workflow diagram}
    \label{fig:workflow}
\end{figure}

Our goal is to systematically benchmark open-source LLMs for automated EMR in Python. We constructed an end-to-end evaluation pipeline that incorporates dataset sampling, preprocessing, prompt generation, LLM inference, automated correctness checks, and metric extraction. The experimental workflow adopted in this study is illustrated in Figure~\ref{fig:workflow}.

\subsection{Pipeline}
As shown in Figure~\ref{fig:workflow}, our experimental pipeline for automated EMR consists of several key stages, supporting two distinct prompting strategies: one-shot and RCI.

Preprocessed Python code and test cases are passed to the Prompt Generator, which builds prompts using the chosen strategy. One-shot uses a single template, while RCI maintains history and iteratively incorporates feedback from prior test failures to refine outputs. The prompt is sent to the LLM inference module, which runs the benchmark models and outputs a candidate refactoring. The Evaluator module checks each candidate's functional correctness by running its test cases for behavioral equivalence. Candidates passing test cases are then analyzed using metrics like CC and LOC. If tests fail, handling depends on the prompting method. One-shot logs the failure, while the RCI appends test feedback to the conversation and retries, up to two attempts. Persistent failures are marked as unsuccessful. The final outputs, including functional correctness, code quality metrics, and, for selected samples, qualitative assessments via a developer survey, form the basis of our comprehensive evaluation.

\subsection{Dataset Selection and Preprocessing}

\subsubsection{Dataset Used}
We use a subset of the CodeNet dataset\footnote{\url{https://github.com/IBM/Project_CodeNet}}\cite{puri2021codenet}, a large-scale corpus of real code submissions collected from competitive programming platforms, containing both code and associated test cases. We used CodeNet because it includes functionally correct solutions validated by problem-specific test cases. Each problem provides the original code and test inputs, allowing reliable post-refactoring correctness checks. While other datasets for refactoring tasks exist, they typically combine multiple types of refactoring, often applied manually or semi-automatically, and may contain inconsistencies or errors that make them less suitable for evaluating EMR specifically. In contrast, CodeNet provides clean, executable, and validated code fragments, enabling us to benchmark Extract Method refactoring in a controlled setting without the confounding factor of noisy or incorrect ground truth. The dataset has been widely used in prior code generation and refactoring studies, supporting reproducibility and comparison.

\subsubsection{Problem and Sample Selection}
From the full set of Python problems (800), we filtered for those that: (a) have a CC of three and sufficient length for EMR. The average CC of the original code files selected is 4.602. (b) Provide at least 40 unique code submissions per problem.

For each of 21 problems, we sampled 40 diverse code submissions (840 total). To reduce redundancy and include both short and long solutions in our benchmark, we analyzed the length distribution (LOC) and oversampled outliers using Laplace smoothing with the following formula:

\[
P_{\text{smoothed}}(x) = \frac{f(x) + \alpha}{N + \alpha \cdot d}
\]

where $f(x)$ represents the frequency of the bin $x$, $N$ is the total number of files, $\alpha$ is the strength (hyperparameter), and $d$ is the number of bins. For our implementation, we set $\alpha = 5$ to apply moderate smoothing, ensuring that rare solution lengths are sufficiently represented without overly distorting the empirical distribution. 

\begin{figure}[h!]
  \centering
  \includegraphics[width=8cm]{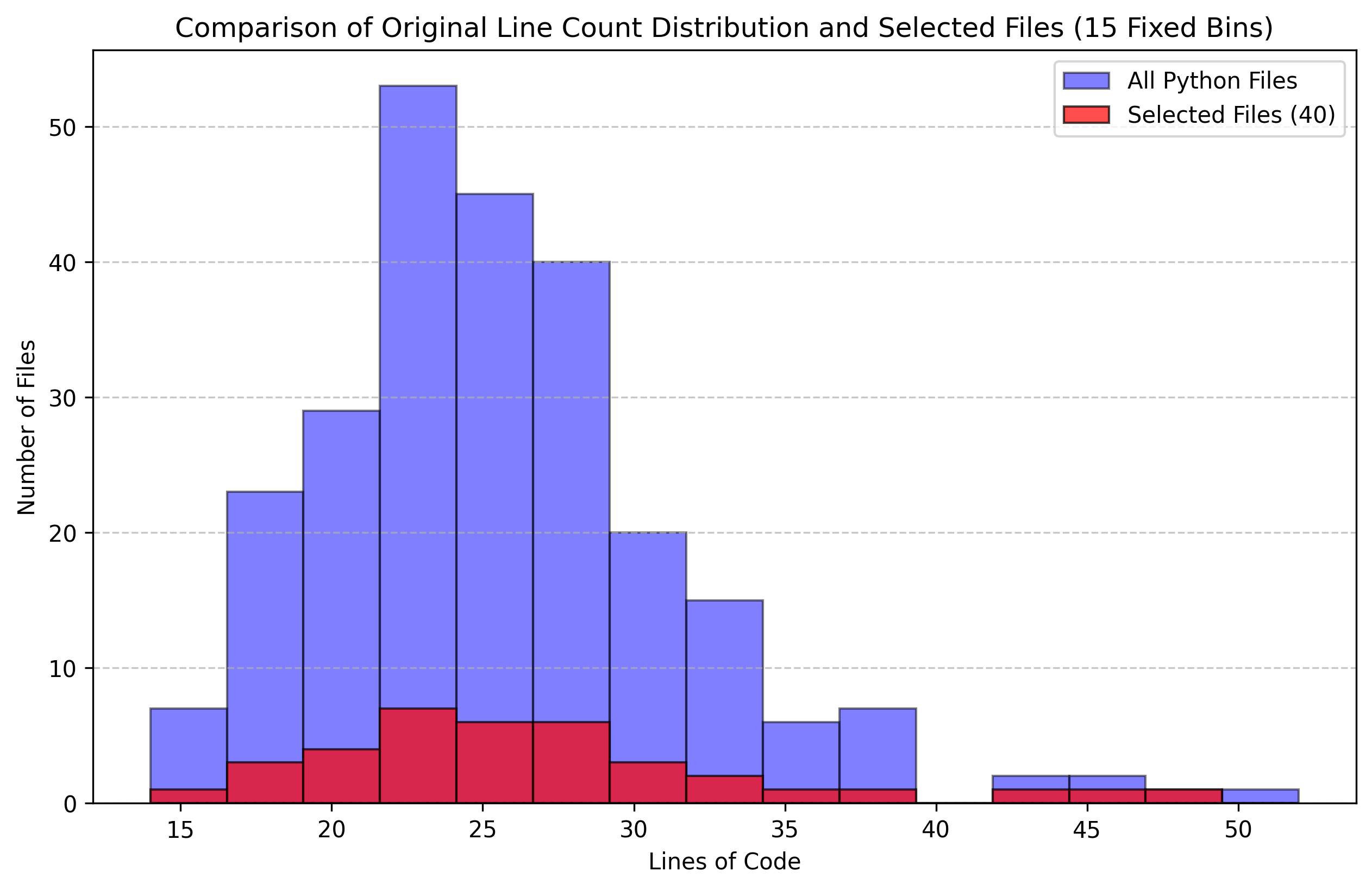}
  \caption{Code sampling distribution of a selected problem}
  \label{fig:loc_dist}
\end{figure}

As shown in Fig. \ref{fig:loc_dist}, the distribution of the selected files was more evenly distributed across different LOC sizes, covering both large and small cases. After this step, we ended up with 40 samples instead of 300 per problem.

\textbf{Preprocessing Steps:}

\begin{enumerate}[label=(\roman*)]
    \item Scope Wrapping: Because Python code often has top-level logic, we wrap global code in a synthetic function (wrapped\_artificially) to enable per-method metric analysis (e.g., CC) while preserving scope. This ensures consistent metrics and fair evaluation.
    \item Deduplication: Samples with identical code after stripping comments and whitespace are deduplicated.
    \item Test Case Validation: Each sample's test cases were extracted and checked on the original code to confirm baseline correctness.
\end{enumerate}
\begin{figure*}[h]
    \centering
    \includegraphics[width=1.0\textwidth]{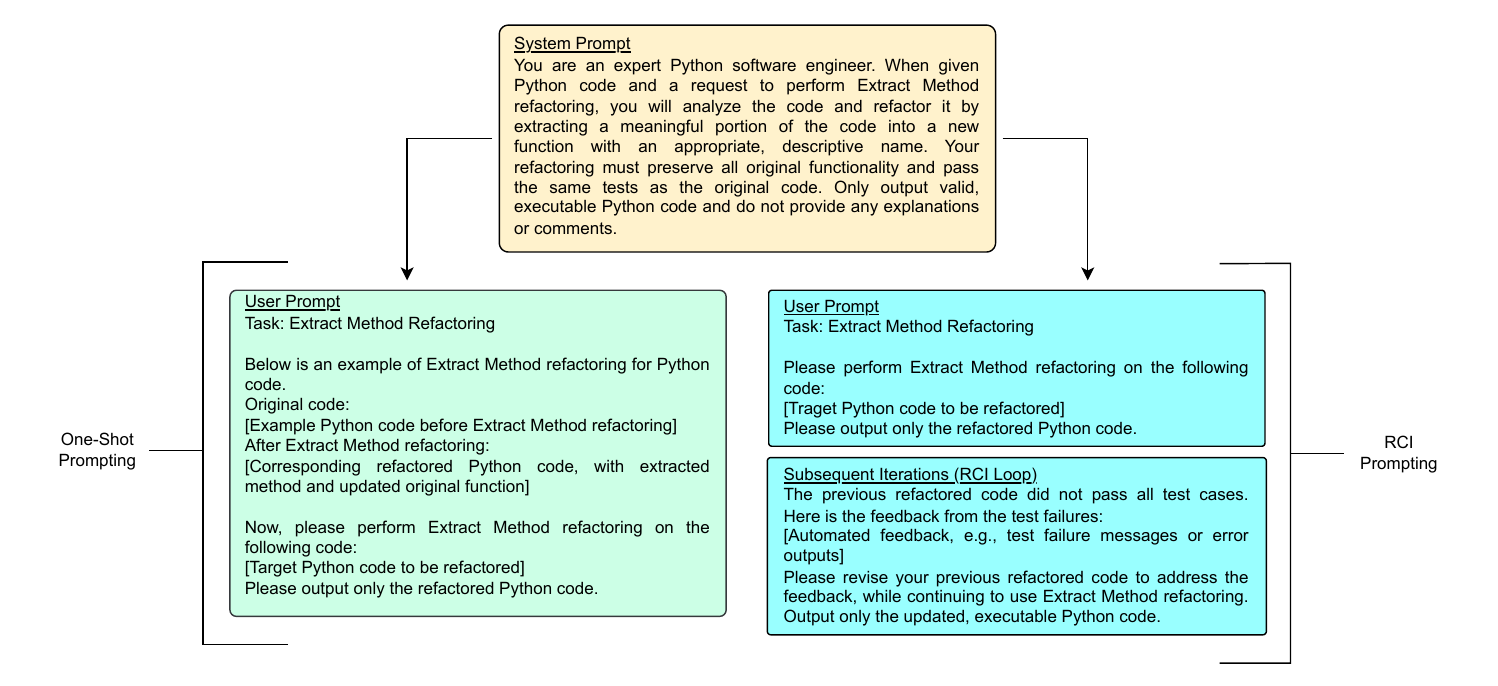}
    \caption{Prompting Templates Used}
    \label{fig:prompt-templates}
\end{figure*}

\subsection{Model Selection}
We selected five state-of-the-art open-source LLMs with manageable compute demands and permissive licensing (Table~\ref{tab:chosen_models}). All models run on a single NVIDIA Tesla V100 GPU (16GB VRAM), reflecting realistic academic and industry budgets and enhancing practical relevance.

\begin{table}[ht]
    \caption{Selected Models}
    \label{tab:chosen_models}
    \centering
    \footnotesize
    \begin{tabular}{|c|c|c|}
        \hline
        \textbf{Model Name} & \textbf{Organization} & \textbf{Param. Size} \\
        \hline
        Qwen/CodeQwen1.5-7B-Chat & Alibaba & 7.2B \\
        \hline
        deepseek-ai/deepseek-coder-6.7b-instruct & DeepSeek & 6.7B \\
        \hline
        meta-llama/Llama-3.2-3B-Instruct & Meta & 3.2B \\
        \hline
        Qwen/Qwen2.5-Coder-7B-Instruct & Alibaba & 7.6B \\
        \hline
        microsoft/Phi-4-mini-instruct & Microsoft & 3.8B \\
        \hline
    \end{tabular}
\end{table}

\subsection{Prompt Design}
This study tests whether advanced prompting, specifically RCI, yields better LLM-generated EMRs than standard one-shot prompting. Both strategies were benchmarked on each code sample.

\begin{enumerate}[label=(\roman*)]
    \item \textbf{One-Shot Prompting:} A single EMR example is provided to guide the model's transformation, minimizing complexity and inference time.
    \item \textbf{Recursive Criticism and Improvement (RCI):} The LLM's refactoring is tested for correctness after each attempt. If incorrect, targeted feedback based on test case errors is added to the prompt, and the model tries again. This process can repeat up to three times, simulating iterative improvement.
\end{enumerate}

Prompt templates were kept consistent across models with minor syntax adjustments (see Figure~\ref{fig:prompt-templates} for examples). Models were instructed to output only executable code, avoid explanations, and limit changes to the EMR.

\subsection{Evaluation}

We use both quantitative and qualitative metrics to assess the EMRs produced by each model and prompting strategy.

\subsubsection{Quantitative Metrics}
The following quantitative metrics are computed for every code sample, model, and prompt type:

\begin{enumerate}[label=\alph*)]
    \item \textbf{Test Pass Percentage}: The proportion of refactored code samples that pass all original test cases, measuring functional correctness. TPP serves as the primary metric for comparing models and prompt strategies head-to-head.
    \item \textbf{Lines of Code}: The average lines per method in the refactored code, reflecting modularity and potential readability improvements.\cite{Morishita2023Refactoring, Hora2020Characteristics, Hummel2016Deriving, Nasrabadi2022An} 
    \item \textbf{Cyclomatic Complexity}: The maximum CC across methods in the refactored code, used to assess the reduction of code complexity.\cite{Alcocer2021Quality, 2023The, Hummel2016Deriving, Nasrabadi2022An}
\end{enumerate}
We evaluate all three metrics for every model and both prompting techniques (one-shot and RCI), providing a direct comparison.
\subsubsection{Qualitative Evaluation}

To complement our quantitative evaluation, we conducted a developer survey to assess the perceived quality and acceptance of EMRs produced by the top-performing models under RCI-based prompting. This survey serves two purposes: (1) to identify which model's refactorings are most favored by human developers, and (2) to investigate whether improvements in automated metrics (TPP, LOC, CC) correspond to higher developer-perceived code quality.

\textbf{\textit{Participants:}} Nine developers with backgrounds in software engineering, data science, and machine learning engineering participated in the survey. The participants varied in their years of professional experience, ranging from 4 years to 8 years.

\textbf{\textit{Survey Design and Scope:}} To keep the survey manageable and avoid participant fatigue, each developer reviewed 20 code samples covering five distinct problems. For each problem, we presented four versions of the code:
\begin{enumerate}[label=\alph*)]
    \item The original (unrefactored) code sample
    \item Three RCI-refactored outputs from the top-performing models based on quantitative metrics (Qwen2.5-Coder, DeepSeek-Coder, and CodeQwen1.5)
\end{enumerate}

This design ensured that all developers assessed the same five problems and the same model outputs for consistent and comparable feedback.

\textbf{\textit{Scoring and Statements:}}
Ratings used a 5-point Likert scale: Totally Disagree (-2), Disagree (-1), Neutral (0), Agree (+1), and Totally Agree (+2). Developers rated each code version according to the following five statements:
\begin{enumerate}[label=\alph*)]
    \item I can easily understand what this code does.
    \item I feel comfortable to add new features to this code.
    \item I can easily troubleshoot the errors and debug this code.
    \item I would accept this refactoring if it was offered by an agent. (refactored versions only)
    \item Even if I do not accept this refactoring, I still find it helpful. (refactored versions only)
\end{enumerate}

\textbf{\textit{Analysis:}}
Ratings were mapped to their numeric weights for each code sample and statement and aggregated across all developers. For each model, a weighted sum provided an overall qualitative score, with positive values indicating generally favorable developer opinions. Acceptance rates were also computed for the statement, "I would accept this refactoring if it was offered by an agent."

The survey took approximately 45 minutes per participant. The results are used to compare qualitative model performance and to analyze the alignment between improvements in quantitative metrics and developer judgments of refactoring quality.

\section{Results}
Our results section presents a comprehensive comparison of five open-source LLMs and two prompting strategies on the EMR task. To structure our analysis, we address the following research questions:

\begin{enumerate}[label=\textbf{RQ\arabic*}]
\item How do different open-source LLMs perform on Extract Method refactoring?
\item How does the prompting strategy (RCI vs. one-shot) affect the quality and correctness of LLM-generated Extract Method refactorings?
\item Are code quality improvements, as measured by automated metrics (LOC, CC), meaningful in practice?
\item Do quantitative metrics align with human developer judgments of good refactoring?
\end{enumerate}

We report both quantitative and qualitative findings to systematically answer these questions, focusing on functional correctness, code quality improvements, developer acceptance, and the relationship between automated metrics and human expectations.

\subsection{Quantitative Results}

To provide an overall comparison of model and prompt performance, Table~\ref{tab:analysis_overall_quantitative_metrics} summarizes the main quantitative metrics: TPP, average LOC per method, and maximum CC, for each approach. This table enables a direct side-by-side assessment of functional correctness and code quality improvements resulting from EMR across all evaluated models and prompting strategies.
\begin{table}[h]
    \centering
    \caption{Quantitative Metrics per Approach Table}
    \renewcommand{\arraystretch}{1.2} % Adjust row height
    \begin{tabular}{|c|c|c|c|c|}
        \hline
        \textbf{Approach} & \textbf{EM TPP} & \textbf{LOC} & \textbf{CC}\\
        
        \hline
        Original Samples & -  & 12.103 & 4.602 \\
        \hline
        CodeQwen-1.5-RCI & 0.638 & 7.337 & 3.668 \\
        \hline
        CodeQwen-1.5-Oneshot & 0.503  & 7.119 & 3.671 \\
        \hline
        Qwen2.5-Coder-RCI & 0.808  & \textbf{5.577} & 3.294 \\
        \hline
        Qwen2.5-Coder-Oneshot & 0.722 & 6.320 & 3.344 \\
        \hline
        Deepseek-Coder-RCI & \textbf{0.829} & 6.192 & 3.453 \\
        \hline
        Deepseek-Coder-Oneshot & 0.743 & 5.827 & 3.411 \\
        \hline
        Llama-3.2-RCI & 0.366 & 8.181 & 2.940 \\
        \hline
        Llama-3.2-Oneshot & 0.336 & 5.801 & \textbf{2.870} \\
        \hline
        Phi-4-RCI & 0.579 & 6.287 & 3.278 \\
        \hline
        Phi-4-Oneshot & 0.462 & 6.652 & 3.317 \\
        \hline
        
    \end{tabular}
    
    \label{tab:analysis_overall_quantitative_metrics}
\end{table}

\textbf{\textit{Test Pass Percentage:}} TPP is our primary metric for assessing the functional correctness of LLM-generated EMRs. TPP is defined as the proportion of generated refactorings for which all test cases pass, ensuring functional equivalence with the original code. This metric is critical: if the refactored code is not functionally correct, improvements in other code quality metrics are largely irrelevant. 

\begin{figure}[h!]
  \centering
  \includegraphics[width=9cm]{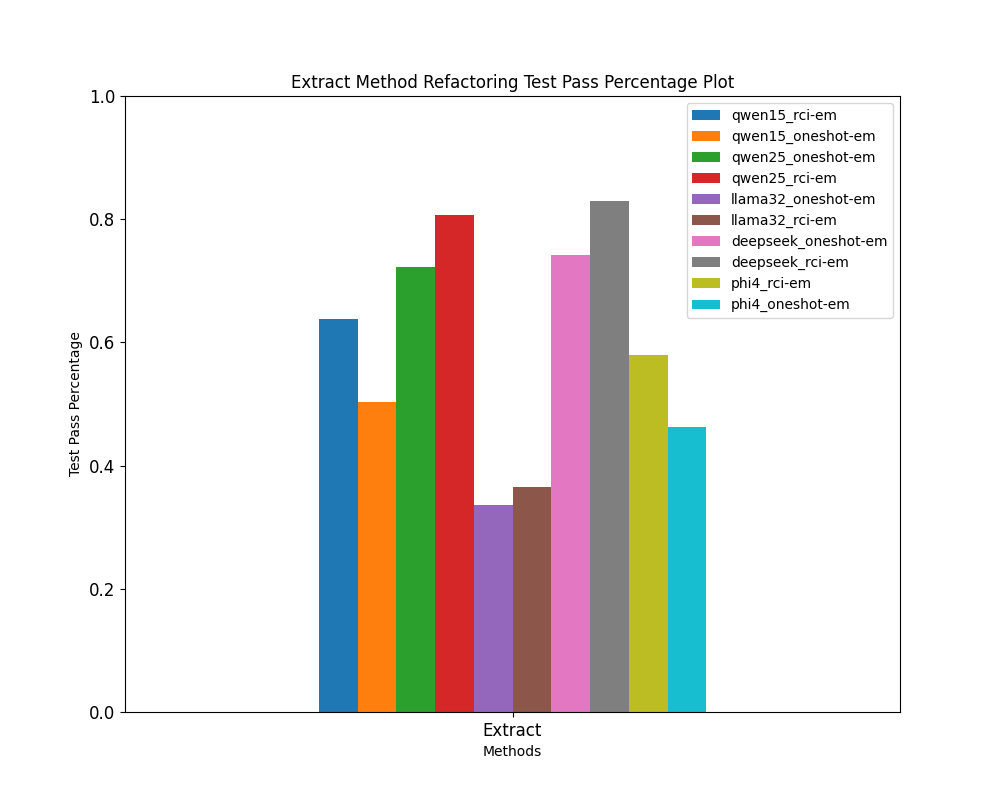}
  \caption{Test Pass Percentage for each Model}
  \label{fig:tpp_bar}
\end{figure}

\begin{figure*}[ht!]
    \centering
    \includegraphics[width=16cm]{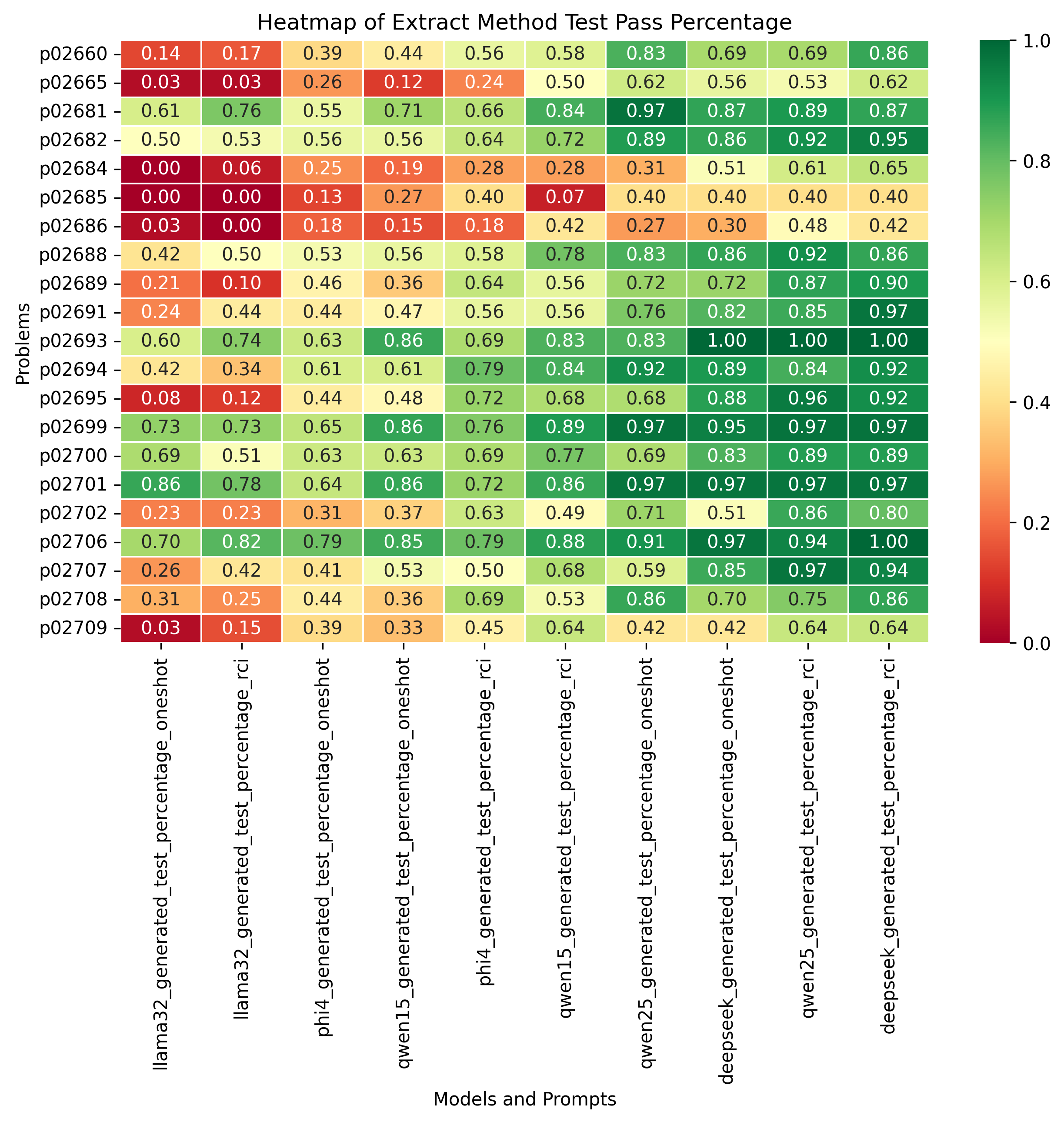}
    \caption{Problem-level Test Pass Heatmap}
    \label{fig:tpp_heatmap}
\end{figure*}

We computed TPP for every model and both prompting strategies (one-shot and RCI) across all 840 code samples. Figure~\ref{fig:tpp_bar} shows the average TPP for each model and prompt type. Figure~\ref{fig:tpp_heatmap} presents a breakdown of TPP for each individual problem and model. This detailed view shows that specific problems, such as p02684, p02685, and p02686, are complex for all models: even the highest-performing models fail to achieve a 50\% pass rate on these tasks.

The results show a clear benefit for RCI prompting: all models achieved higher TPP than one-shot, confirming the value of iterative feedback for improving LLM reliability. Gains were largest for high-performing models. \textbf{Qwen2.5-Coder} and \textbf{DeepSeek-Coder} led with RCI TPPs of 80.8\% and 82.9\%, up from One-shot's 72.2\% and 74.3\%, outperforming all others. \textbf{CodeQwen1.5} improved from 50.3\% to 63.8\%, showing mid-tier models also benefit. The smallest models (\textbf{Phi-4}, \textbf{Llama3.2}) remained below 60\% TPP, with smaller absolute gains, likely due to limited capacity.

To better understand the effect of iterative prompting, we analyzed the distribution of successful refactorings across RCI iterations for each model. We observed that most improvements occurred early: for instance, Qwen2.5-Coder and DeepSeek-Coder, successfully refactored about 64\% and 67\% of cases, respectively, in the first iteration, with a further 25\% and 30\% corrected after the second. Importantly, even with this simple capped budget, RCI delivered $>$10 percentage point improvements in TPP, showing robustness across models. By design, initial attempt of RCI is a zero-shot attempt, the model is given only the task specification and target code (no examples, no test-failure feedback). Zero-shot results were consistently lower than both one-shot and RCI conditions, TPP = 0.472 for CodeQwen-1.5, 0.686 for Qwen2.5-Coder, 0.706 for DeepSeek-Coder, 0.334 for Llama-3.2, and 0.427 for Phi-4. Iterative feedback in subsequent RCI rounds raised these to 0.638, 0.808, 0.829, 0.366, and 0.579, respectively, confirming that most of the observed gains originate from the feedback-driven iterations rather than the initial attempt.

\begin{mdframed}[style=graybox]
\footnotesize
\textbf{TPP Summary:} RCI-based prompting consistently outperforms one-shot prompting in functional correctness (TPP) across all models. The highest reliability is achieved by Qwen2.5-Coder and DeepSeek-Coder, while smaller models like Llama3.2 and Phi-4 lag behind, highlighting both the importance of prompt strategy and model selection for effective EMR.
\end{mdframed} 

\begin{figure*}[!hbt]
    \centering
    \begin{subfigure}[b]{0.49\textwidth}
        \centering
        \includegraphics[width=\textwidth]{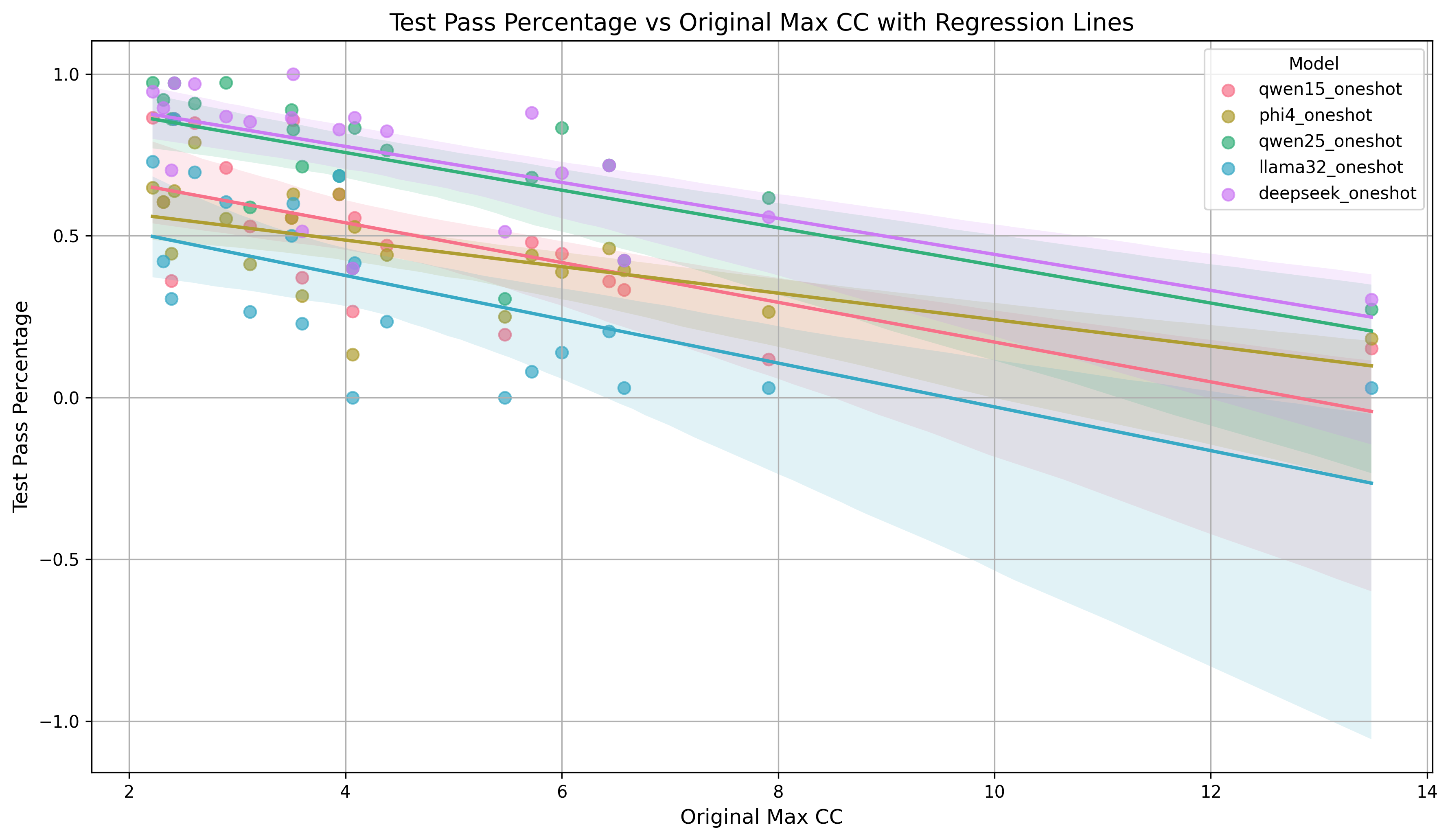}
        \caption{Regression Lines for TPP versus CC}
        \label{fig:tpp_vs_cc_regression}
    \end{subfigure}
    \begin{subfigure}[b]{0.49\textwidth}
    \centering
    \includegraphics[width=\textwidth]{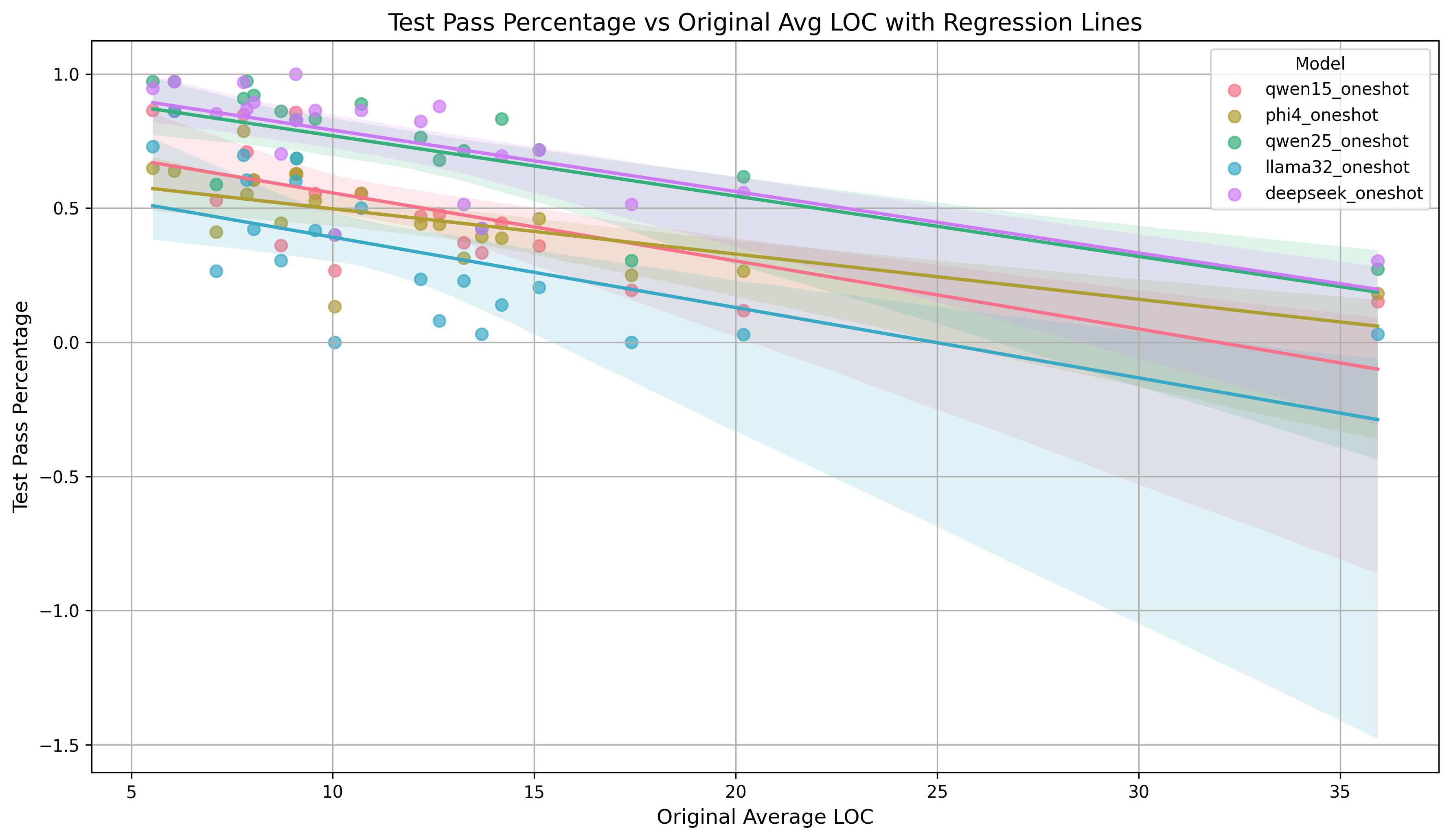}
    \caption{Regression Lines for TPP versus LOC}
    \label{fig:tpp_vs_loc_regression}
    \end{subfigure}
    \caption{Regression Lines for TPP vs Quantitative metrics used }
        \label{fig:tpp_vs_regression}
\end{figure*}

\textbf{\textit{Lines of Code:}} LOC per method serves as an indicator of the modularity achieved by EMR. As reported in Table~\ref{tab:analysis_overall_quantitative_metrics}, the original samples have an average LOC per method of 12.1. The best-performing approaches: Qwen2.5-Coder-RCI (5.58) and DeepSeek-Coder-RCI (6.19) achieve more than a 50\% reduction in LOC compared to the original code. Other models, such as CodeQwen1.5-RCI and Phi-4-RCI, also demonstrate strong reductions, while Llama3.2 achieves a more modest improvement.

However, the relation between LOC and functional correctness is non-trivial. As visualized in Figure~\ref{fig:tpp_vs_loc_regression}, which presents regression lines for TPP versus LOC for each model, there is a clear trend: models with higher TPP also achieve lower LOC, indicating that reliable refactoring is correlated with improved modularity. Yet, some models (notably Llama3.2) report low LOC values while also having poor TPP. This is due to incomplete or trivial code generations that may be short, but do not produce functionally correct solutions. Therefore, as emphasized in the thesis, LOC should always be interpreted alongside TPP: substantial reductions in LOC are only meaningful if the code remains correct and passes all test cases.

The results highlight that the most effective EMR, as achieved by Qwen2.5-Coder and DeepSeek-Coder, balances both high functional correctness and substantial modularization of code.

\begin{mdframed}[style=graybox]
\footnotesize
\textbf{LOC Summary:} Effective EMR with top-performing models reduces LOC per method by over 50\%, and significant LOC improvements are only meaningful when accompanied by high functional correctness.
\end{mdframed}

\textbf{\textit{Cyclomatic Complexity:}} CC measures the logical complexity of code, with lower values indicating simpler and potentially more maintainable functions. As shown in Table~\ref{tab:analysis_overall_quantitative_metrics}, the original code samples have an average maximum CC of 4.60. All models reduce this metric after EMR, with Qwen2.5-Coder-RCI and DeepSeek-Coder-RCI achieving strong results at 3.29 and 3.45, respectively. Interestingly, Llama3.2-Oneshot and Phi-4-RCI report the lowest CC values (2.87 and 3.28), but as with LOC, these models also produce many incomplete or incorrect outputs, as indicated by their low TPP.

This nuanced relationship is visualized in Figure~\ref{fig:tpp_vs_cc_regression}, which presents regression lines for TPP versus CC for each model. The plot shows that the most meaningful reductions in CC are achieved by models that also maintain high functional correctness. Models with very low CC but poor TPP, like Llama3.2, may achieve simplicity by oversimplifying code or omitting logic, rather than through proper refactoring.

\begin{mdframed}[style=graybox]
\footnotesize
\textbf{CC Summary:} CC and TPP should always be considered together, and the best-performing model (Qwen2.5-Coder-RCI) reduced average maximum CC from 4.60 to 3.29 while maintaining the highest TPP.
\end{mdframed}

\subsection{Qualitative Evaluation}
\begin{figure}[hbt!]
    \centering
    \includegraphics[width=9cm]{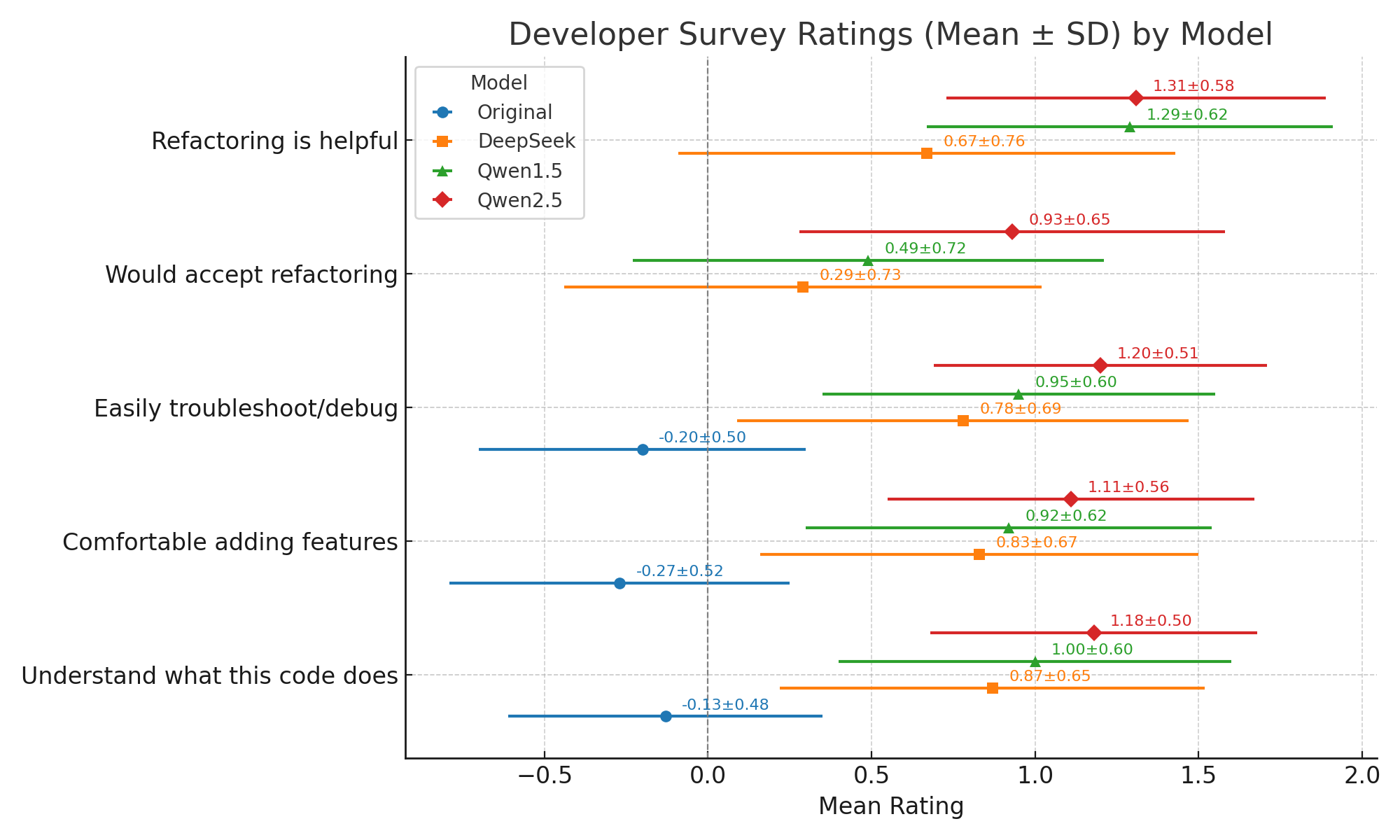}
    \captionsetup{justification=centering}
    \caption{Mean (Standard Deviation) of Developer Survey Ratings per Model and Statement.}
    \caption*{\small Likert scale: $-2$ = Totally Disagree, $0$ = Neutral, $+2$ = Totally Agree.}
    \label{fig:survey}
\end{figure}
To capture human-centric perspectives on code quality, we surveyed developers to assess the refactoring outputs of the top three RCI-prompted models, analyzing perceptions of Readability, Maintainability, and overall acceptability.

%\begin{table*}[h]
%    \centering
%    \caption{Mean (Standard Deviation) of Developer Survey Ratings per Model and Statement. Likert scale: $-2$ = Totally Disagree, $0$ = Neutral, $+2$ = Totally Agree.}
%    \renewcommand{\arraystretch}{1.2} % Adjust row height
%    \begin{tabular}{|p{7.5cm}|c|c|c|c|}
%        \hline
%        \textbf{Statement} & \textbf{Original} & \textbf{DeepSeek} & \textbf{Qwen1.5} & \textbf{Qwen2.5} \\
%        \hline
%        I can easily understand what this code does. & $-0.13\ (0.48)$ & $0.87\ (0.65)$ & $1.00\ (0.60)$ & $1.18\ (0.50)$ \\
%        \hline
%        I feel comfortable to add new features into this code. & $-0.27\ (0.52)$ & $0.83\ (0.67)$ & $0.92\ (0.62)$ & $1.11\ (0.56)$ \\
%        \hline
%        I can easily troubleshoot the errors and debug this code. & $-0.20\ (0.50)$ & $0.78\ (0.69)$ & $0.95\ (0.60)$ & $1.20\ (0.51)$ \\
%        \hline
%        I would accept this refactoring if it was offered by an agent. & -- & $0.29\ (0.73)$ & $0.49\ (0.72)$ & $0.93\ (0.65)$ \\
%        \hline
%        Even if I do not accept this refactoring, I still find it helpful. & -- & $0.67\ (0.76)$ & $1.29\ (0.62)$ & $1.31\ (0.58)$ \\
%        \hline
%    \end{tabular}
%    \label{tab:survey_meansd}
%\end{table*}

Nine developers with varied backgrounds and experience levels participated in the survey, reviewing 20 code samples each. For every sample, developers rated the original code and three RCI-refactored versions. Figure \ref{fig:survey} reports the mean and standard deviation of developer ratings for each statement and model, on a five-point Likert scale (from $-2 =$ Totally Disagree to $+2 =$ Totally Agree). To assess the consistency of human judgments, we computed the quadratic-weighted Cohen’s $\kappa$ across all rater pairs, obtaining an average $\kappa \approx 0.30$, with individual values ranging from 0.17 to 0.52 depending on the survey statement. According to the Landis \& Koch scale, these levels correspond to fair-to-moderate agreement. At the model level, $\kappa$ ranged from 0.24 for Qwen2.5-Coder to 0.34 for CodeQwen-1.5. These results indicate that, despite some natural subjectivity, developer ratings were sufficiently consistent to support our conclusions.

Results show that Qwen2.5-Coder achieves the highest average scores for all statements, with relatively low standard deviations indicating strong agreement among developers. In contrast, the original code was consistently rated below neutral, especially in statements related to Readability and Maintainability.

\textbf{\textit{Readability :}} The statement "I can easily understand what this code does" directly measures the perceived clarity of the code. Qwen2.5-Coder achieved the highest average readability score ($1.18 \pm 0.50$), closely followed by Qwen1.5 ($1.00 \pm 0.60$). DeepSeek-Coder also performed positively ($0.87 \pm 0.65$), while the original code was rated slightly below neutral ($-0.13 \pm 0.48$). This demonstrates that when properly prompted, all top LLMs produce refactorings that are easier for developers to understand than the unrefactored code.

\textbf{\textit{Maintainability :}} For Maintainability, the statements "I feel comfortable adding new features into this code" and "I can easily troubleshoot the errors and debug this code" were both consistently rated higher for LLM-generated refactorings than for the originals. Qwen2.5-Coder again led with means above $1.1$ for both statements, while the original code lagged, with average scores below zero.

\textbf{\textit{Acceptance and Practical Helpfulness:}} The strictest test was the statement "I would accept this refactoring if it was offered by an agent." Here, Qwen2.5-Coder still scored highest ($0.93 \pm 0.65$), though average ratings were lower than for the readability and maintainability statements, highlighting the more cautious stance developers may take toward fully automated changes. Qwen1.5 and DeepSeek-Coder followed at $0.49$ and $0.29$, respectively. Notably, all refactored outputs were viewed as improvements over the original, as shown by the generally positive mean values. Additionally, the more forgiving statement "Even if I do not accept this refactoring, I still find it helpful" saw the highest scores overall, especially for Qwen2.5-Coder ($1.31 \pm 0.58$) and Qwen1.5 ($1.29 \pm 0.62$). This suggests that even when developers are not ready to directly accept an LLM-generated refactoring, they still find its output valuable as a starting point or inspiration for their own edits.
    
\textbf{\textit{Consistency and Variability:}} Standard deviations for all models and statements were generally moderate (typically in the range of 0.5–0.7), indicating reasonable agreement among participants. Slightly higher variability was observed for the more subjective acceptance questions, reflecting natural diversity in developer trust and style.

\begin{mdframed}[style=graybox]
\footnotesize
\textbf{Qualitative Evaluation Summary:} Qwen2.5-Coder’s RCI refactorings scored highest in Readability, Maintainability, and acceptance, outperforming both the original code and other models. Overall, model refactorings were rated above original submissions, confirming the qualitative benefits of LLM-based EMR.
\end{mdframed}

\subsection{Alignment Between Quantitative and Qualitative Results}

A key research question in this study is whether improvements in quantitative metrics, specifically TPP, LOC, and CC, reliably reflect the code quality and acceptability perceived by human developers.

For Qwen2.5-Coder, the relationship is clear and positive. It achieved the highest TPP (0.808), the lowest average LOC per method (5.58), and one of the lowest maximum CC values (3.29) among all tested approaches, as shown in  Table \ref{tab:analysis_overall_quantitative_metrics}. Correspondingly, it received the highest mean developer ratings across all qualitative survey statements, with an average score of $1.18 \pm $0.50 for Readability and $0.93 \pm $0.65 for acceptance ("I would accept this refactoring if it was offered by an agent.").

However, for other models, the alignment is less straightforward. DeepSeek-Coder slightly outperformed Qwen2.5-Coder in TPP (0.829 vs. 0.808) and had comparably low LOC and CC (6.19 and 3.45, respectively), but its mean acceptance score in the survey was only $0.29 \pm $0.73, noticeably lower than both Qwen2.5-Coder and Qwen1.5  ($0.49 \pm $0.72).

Qwen1.5, although having lower TPP (0.638) and higher LOC (7.34) and CC (3.67), was more favorably rated by developers than DeepSeek-Coder. For instance, it scored $1.00 \pm $0.60 for Readability and $0.49 \pm $0.72 for acceptance, second only to Qwen2.5-Coder.

\begin{mdframed}[style=graybox]
\footnotesize
\textbf{Alignment Summary:} While quantitative metrics provide valuable signals, they are insufficient substitutes for developer-centered evaluation and can be misleading if used in isolation.
\end{mdframed}

\section{Discussion}

Our systematic evaluation of five open-source LLMs for EMR demonstrates that model choice and prompting strategy significantly affect both functional correctness and code quality.

\textbf{\textit{LLMs' performance on EMR:}} Our results reveal a varied level of performance among open-source LLMs on the EMR task. Qwen2.5-Coder and DeepSeek-Coder achieved functional correctness (TPP) above 80\% and consistently produced more modular and less complex code. In contrast, smaller models such as Llama3.2 and Phi-4 exhibited substantially lower TPP and often failed to generate correct refactorings, highlighting the critical importance of model selection in practical automated refactoring scenarios.

\textbf{\textit{Impact of Prompting Strategy:}} Prompting strategy had a significant impact on both functional correctness and code quality. Across all models, RCI-based prompting consistently outperformed one-shot prompting, with average TPP improvements of 8–14 percentage points depending on the model. RCI prompting enabled iterative feedback and correction, resulting in more reliable and robust refactorings. These findings underscore that prompt engineering is a key factor in unlocking LLM performance for software engineering tasks.

\textbf{\textit{Are code quality improvements meaningful:}}
Automated metrics such as LOC and CC provide useful indications of code modularity and complexity, but our results show they are only meaningful when functional correctness is preserved. The best models achieved both high TPP and substantial reductions in LOC and CC, but some models achieved low complexity or short code at the expense of correctness, producing results that would not be acceptable to developers. This demonstrates that automated code quality metrics must always be interpreted in conjunction with functional validation.

\textbf{\textit{Quantitative metrics-human judgements allignment:}} Our developer survey shows that top models, notably \textbf{Qwen2.5-Coder}, paired high quantitative scores with strong developer acceptance. In contrast, models like \textbf{DeepSeek-Coder} achieved strong quantitative results but lower developer preference. This misalignment highlights the need to combine human feedback with automated metrics when evaluating LLM-driven refactoring.

\section{Threat to Validity}
We acknowledge several potential threats and limitations that might hinder the results achieved in this study: 

\textbf{\textit{Internal Validity:}} The RCI prompting approach uses automated test failures as feedback for language models, but this may overlook nuanced or contextual guidance. While the feedback loop is standardized to reduce bias and ensure consistency, some error types may not be effectively addressed. Future work should consider richer feedback, such as static analysis, human critique, or natural language explanations.

\textbf{\textit{External Validity:}} Our evaluation focuses on Python code from the CodeNet dataset and uses only EMR. Since language features vary, results may not generalize to other languages or refactoring types. And the programs analyzed in this study are relatively small, with a median size of approximately 30 lines of code, as shown in Figure \ref{fig:loc_dist}. This limited size may affect the generalizability of the results to larger, more complex programs. To address this, we selected diverse problems from CodeNet covering a range of styles and complexities. The human study, though carefully designed, involved nine developers and a fixed set of problems, limiting generalizability. Participants had varied backgrounds and all assessed the same problems to ensure comparability. Future work should scale up with more developers, tasks, and languages.

\textbf{\textit{Construct Validity:}} We use widely accepted automated metrics such as TPP, LOC, and CC, alongside a standard Likert-scale developer survey to assess code quality and acceptance. Survey statements were pilot-tested and refined for clarity. However, it is possible that our chosen metrics and survey items do not fully capture some important aspects of code quality or developer preference.

\textbf{\textit{Reproducibility:}} To facilitate replication and extension, we will release our full evaluation pipeline, code, data splits, and model configurations upon publication. This will allow other researchers to apply our methodology to additional datasets, languages, or refactoring tasks.

\section{Future Works}
Our study opens up several promising directions for further research and improvement.
While this work focuses on EMR in Python, future studies should apply the benchmarking framework to additional programming languages (such as Java, C++, or JavaScript) and to a broader range of refactoring operations. This would allow for more general conclusions and reveal whether the strengths and weaknesses observed in Python transfer to other language paradigms and code structures.

Scaling up the developer survey, both in participant number and problem diversity, would yield deeper insights into how LLM-generated refactorings are perceived in practice. Incorporating qualitative interviews, open-ended developer feedback, and longitudinal studies on code maintainability after automated refactoring could further enrich our understanding of real-world impact.

Future work could explore more sophisticated forms of feedback in RCI prompting, such as static analysis hints, natural language critiques, or brief human-in-the-loop interventions. Evaluating how richer feedback improves LLM performance or acceptance could lead to more effective and trusted refactoring agents.

Integrating LLM-based refactoring tools into development environments (e.g., IDEs, CI/CD pipelines) and monitoring their impact in live software projects provides evidence of practical benefits and potential challenges. This could include tracking developer acceptance rates, post-refactoring defect rates, or productivity gains over time.

Continued expansion of code and refactoring benchmarks, including more diverse codebases, more realistic enterprise code, and additional test suites, would further validate the scalability and robustness of both LLM and traditional approaches.

\section{Conclusion}
This paper evaluates open-source large language models for automated EMR in Python, comparing five state-of-the-art models and two prompting strategies. Our results demonstrate that model selection and prompt engineering critically affect functional correctness and code quality, with recursive criticism and improvement (RCI) consistently outperforming standard one-shot prompting. The best-performing models, particularly Qwen2.5-Coder and DeepSeek-Coder, achieve high test pass rates and meaningful code complexity and modularity metrics reductions.
Our qualitative developer survey reveals that improvements in quantitative metrics do not always align with developer acceptance or perceived code quality, emphasizing the importance of human-centered evaluation in automated refactoring research. Our findings show that when properly guided, open-source LLMs can now produce automated refactorings that are both reliable and acceptable to developers.
As automated code refactoring powered by LLMs matures, integrating developer feedback and thoughtful, prompt design will be key to building practical, trustworthy tools for real-world software engineering.

\bibliographystyle{IEEEtran}
\bibliography{conference_101719}

\end{document}